\documentclass[conference,10pt,twoside,twocolumn]{IEEEtran}
\IEEEoverridecommandlockouts
\usepackage{cite}
\usepackage[T1]{fontenc}
\usepackage{graphicx}
\usepackage{amssymb}
\usepackage{amsmath}
\usepackage{amsthm}
\usepackage{subfigure}
\usepackage{booktabs} 
\usepackage{multirow}
\usepackage{microtype}
\usepackage{balance}
\usepackage{xcolor}

\usepackage[hidelinks]{hyperref}

\usepackage{amssymb}
\usepackage{amsfonts}
\usepackage{mathrsfs}
\usepackage{xspace}
\usepackage{bm}
\usepackage{upgreek}

\newcommand{\safemath}[2]{\newcommand{#1}{\ensuremath{#2}\xspace}}

\safemath{\bma}{\mathbf{a}}
\safemath{\bmb}{\mathbf{b}}
\safemath{\bmc}{\mathbf{c}}
\safemath{\bmd}{\mathbf{d}}
\safemath{\bme}{\mathbf{e}}
\safemath{\bmf}{\mathbf{f}}
\safemath{\bmg}{\mathbf{g}}
\safemath{\bmh}{\mathbf{h}}
\safemath{\bmi}{\mathbf{i}}
\safemath{\bmj}{\mathbf{j}}
\safemath{\bmk}{\mathbf{k}}
\safemath{\bml}{\mathbf{l}}
\safemath{\bmm}{\mathbf{m}}
\safemath{\bmn}{\mathbf{n}}
\safemath{\bmo}{\mathbf{o}}
\safemath{\bmp}{\mathbf{p}}
\safemath{\bmq}{\mathbf{q}}
\safemath{\bmr}{\mathbf{r}}
\safemath{\bms}{\mathbf{s}}
\safemath{\bmt}{\mathbf{t}}
\safemath{\bmu}{\mathbf{u}}
\safemath{\bmv}{\mathbf{v}}
\safemath{\bmw}{\mathbf{w}}
\safemath{\bmx}{\mathbf{x}}
\safemath{\bmy}{\mathbf{y}}
\safemath{\bmz}{\mathbf{z}}
\safemath{\bmzero}{\mathbf{0}}
\safemath{\bmone}{\mathbf{1}}

\bmdefine{\biad}{a}
\bmdefine{\bibd}{b}
\bmdefine{\bicd}{c}
\bmdefine{\bidd}{d}
\bmdefine{\bied}{e}
\bmdefine{\bifd}{f}
\bmdefine{\bigd}{g}
\bmdefine{\bihd}{h}
\bmdefine{\biid}{i}
\bmdefine{\bijd}{j}
\bmdefine{\bikd}{k}
\bmdefine{\bild}{l}
\bmdefine{\bimd}{m}
\bmdefine{\bind}{n}
\bmdefine{\biod}{o}
\bmdefine{\bipd}{p}
\bmdefine{\biqd}{q}
\bmdefine{\bird}{r}
\bmdefine{\bisd}{s}
\bmdefine{\bitd}{t}
\bmdefine{\biud}{u}
\bmdefine{\bivd}{v}
\bmdefine{\biwd}{w}
\bmdefine{\bixd}{x}
\bmdefine{\biyd}{y}
\bmdefine{\bizd}{z}

\bmdefine{\bixid}{\xi}
\bmdefine{\bilambdad}{\lambda}
\bmdefine{\bimud}{\mu}
\bmdefine{\bithetad}{\theta}
\bmdefine{\biphid}{\phi}
\bmdefine{\bideltad}{\delta}

\safemath{\bmia}{\biad}
\safemath{\bmib}{\bibd}
\safemath{\bmic}{\bicd}
\safemath{\bmid}{\bidd}
\safemath{\bmie}{\bied}
\safemath{\bmif}{\bifd}
\safemath{\bmig}{\bigd}
\safemath{\bmih}{\bihd}
\safemath{\bmii}{\biid}
\safemath{\bmij}{\bijd}
\safemath{\bmik}{\bikd}
\safemath{\bmil}{\bild}
\safemath{\bmim}{\bimd}
\safemath{\bmin}{\bind}
\safemath{\bmio}{\biod}
\safemath{\bmip}{\bipd}
\safemath{\bmiq}{\biqd}
\safemath{\bmir}{\bird}
\safemath{\bmis}{\bisd}
\safemath{\bmit}{\bitd}
\safemath{\bmiu}{\biud}
\safemath{\bmiv}{\bivd}
\safemath{\bmiw}{\biwd}
\safemath{\bmix}{\bixd}
\safemath{\bmiy}{\biyd}
\safemath{\bmiz}{\bizd}

\safemath{\bmxi}{\bixid}
\safemath{\bmlambda}{\bilambdad}
\safemath{\bmmu}{\bimud}
\safemath{\bmtheta}{\bithetad}
\safemath{\bmphi}{\biphid}
\safemath{\bmdelta}{\bideltad}

\safemath{\bA}{\mathbf{A}}
\safemath{\bB}{\mathbf{B}}
\safemath{\bC}{\mathbf{C}}
\safemath{\bD}{\mathbf{D}}
\safemath{\bE}{\mathbf{E}}
\safemath{\bF}{\mathbf{F}}
\safemath{\bG}{\mathbf{G}}
\safemath{\bH}{\mathbf{H}}
\safemath{\bI}{\mathbf{I}}
\safemath{\bJ}{\mathbf{J}}
\safemath{\bK}{\mathbf{K}}
\safemath{\bL}{\mathbf{L}}
\safemath{\bM}{\mathbf{M}}
\safemath{\bN}{\mathbf{N}}
\safemath{\bO}{\mathbf{O}}
\safemath{\bP}{\mathbf{P}}
\safemath{\bQ}{\mathbf{Q}}
\safemath{\bR}{\mathbf{R}}
\safemath{\bS}{\mathbf{S}}
\safemath{\bT}{\mathbf{T}}
\safemath{\bU}{\mathbf{U}}
\safemath{\bV}{\mathbf{V}}
\safemath{\bW}{\mathbf{W}}
\safemath{\bX}{\mathbf{X}}
\safemath{\bY}{\mathbf{Y}}
\safemath{\bZ}{\mathbf{Z}}

\safemath{\bZero}{\mathbf{0}}
\safemath{\bOne}{\mathbf{1}}
\safemath{\bDelta}{\mathbf{\Delta}}
\safemath{\bLambda}{\mathbf{\Lambda}}
\safemath{\bPhi}{\mathbf{\Upphi}}
\safemath{\bSigma}{\mathbf{\Upsigma}}
\safemath{\bOmega}{\mathbf{\Upomega}}
\safemath{\bTheta}{\mathbf{\Uptheta}}

\bmdefine{\biAd}{A}
\bmdefine{\biBd}{B}
\bmdefine{\biCd}{C}
\bmdefine{\biDd}{D}
\bmdefine{\biEd}{E}
\bmdefine{\biFd}{F}
\bmdefine{\biGd}{G}
\bmdefine{\biHd}{H}
\bmdefine{\biId}{I}
\bmdefine{\biJd}{J}
\bmdefine{\biKd}{K}
\bmdefine{\biLd}{L}
\bmdefine{\biMd}{M}
\bmdefine{\biOd}{N}
\bmdefine{\biPd}{O}
\bmdefine{\biQd}{P}
\bmdefine{\biRd}{R}
\bmdefine{\biSd}{S}
\bmdefine{\biTd}{T}
\bmdefine{\biUd}{U}
\bmdefine{\biVd}{V}
\bmdefine{\biWd}{W}
\bmdefine{\biXd}{X}
\bmdefine{\biYd}{Y}
\bmdefine{\biZd}{Z}

\bmdefine{\biDelta}{\Delta}
\bmdefine{\biLambda}{\Lambda}
\bmdefine{\biPhi}{\Phi}
\bmdefine{\biSigma}{\Sigma}
\bmdefine{\biOmega}{\Omega}
\bmdefine{\biTheta}{\Theta}

\safemath{\bimA}{\biAd}
\safemath{\bimB}{\biBd}
\safemath{\bimC}{\biCd}
\safemath{\bimD}{\biDd}
\safemath{\bimE}{\biEd}
\safemath{\bimF}{\biFd}
\safemath{\bimG}{\biGd}
\safemath{\bimH}{\biHd}
\safemath{\bimI}{\biId}
\safemath{\bimJ}{\biJd}
\safemath{\bimK}{\biKd}
\safemath{\bimL}{\biLd}
\safemath{\bimM}{\biMd}
\safemath{\bimN}{\biNd}
\safemath{\bimO}{\biOd}
\safemath{\bimP}{\biPd}
\safemath{\bimQ}{\biQd}
\safemath{\bimR}{\biRd}
\safemath{\bimS}{\biSd}
\safemath{\bimT}{\biTd}
\safemath{\bimU}{\biUd}
\safemath{\bimV}{\biVd}
\safemath{\bimW}{\biWd}
\safemath{\bimX}{\biXd}
\safemath{\bimY}{\biYd}
\safemath{\bimZ}{\biZd}

\safemath{\bimDelta}{\biDelta}
\safemath{\bimLambda}{\biLambda}
\safemath{\bimPhi}{\biPhi}
\safemath{\bimSigma}{\biSigma}
\safemath{\bimOmega}{\biOmega}
\safemath{\bimTheta}{\biTheta}

\safemath{\setA}{\mathcal{A}}
\safemath{\setB}{\mathcal{B}}
\safemath{\setC}{\mathcal{C}}
\safemath{\setD}{\mathcal{D}}
\safemath{\setE}{\mathcal{E}}
\safemath{\setF}{\mathcal{F}}
\safemath{\setG}{\mathcal{G}}
\safemath{\setH}{\mathcal{H}}
\safemath{\setI}{\mathcal{I}}
\safemath{\setJ}{\mathcal{J}}
\safemath{\setK}{\mathcal{K}}
\safemath{\setL}{\mathcal{L}}
\safemath{\setM}{\mathcal{M}}
\safemath{\setN}{\mathcal{N}}
\safemath{\setO}{\mathcal{O}}
\safemath{\setP}{\mathcal{P}}
\safemath{\setQ}{\mathcal{Q}}
\safemath{\setR}{\mathcal{R}}
\safemath{\setS}{\mathcal{S}}
\safemath{\setT}{\mathcal{T}}
\safemath{\setU}{\mathcal{U}}
\safemath{\setV}{\mathcal{V}}
\safemath{\setW}{\mathcal{W}}
\safemath{\setX}{\mathcal{X}}
\safemath{\setY}{\mathcal{Y}}
\safemath{\setZ}{\mathcal{Z}}
\safemath{\emptySet}{\varnothing}

\safemath{\colA}{\mathscr{A}}
\safemath{\colB}{\mathscr{B}}
\safemath{\colC}{\mathscr{C}}
\safemath{\colD}{\mathscr{D}}
\safemath{\colE}{\mathscr{E}}
\safemath{\colF}{\mathscr{F}}
\safemath{\colG}{\mathscr{G}}
\safemath{\colH}{\mathscr{H}}
\safemath{\colI}{\mathscr{I}}
\safemath{\colJ}{\mathscr{J}}
\safemath{\colK}{\mathscr{K}}
\safemath{\colL}{\mathscr{L}}
\safemath{\colM}{\mathscr{M}}
\safemath{\colN}{\mathscr{N}}
\safemath{\colO}{\mathscr{O}}
\safemath{\colP}{\mathscr{P}}
\safemath{\colQ}{\mathscr{Q}}
\safemath{\colR}{\mathscr{R}}
\safemath{\colS}{\mathscr{S}}
\safemath{\colT}{\mathscr{T}}
\safemath{\colU}{\mathscr{U}}
\safemath{\colV}{\mathscr{V}}
\safemath{\colW}{\mathscr{W}}
\safemath{\colX}{\mathscr{X}}
\safemath{\colY}{\mathscr{Y}}
\safemath{\colZ}{\mathscr{Z}}

\safemath{\opA}{\mathbb{A}}
\safemath{\opB}{\mathbb{B}}
\safemath{\opC}{\mathbb{C}}
\safemath{\opD}{\mathbb{D}}
\safemath{\opE}{\mathbb{E}}
\safemath{\opF}{\mathbb{F}}
\safemath{\opG}{\mathbb{G}}
\safemath{\opH}{\mathbb{H}}
\safemath{\opI}{\mathbb{I}}
\safemath{\opJ}{\mathbb{J}}
\safemath{\opK}{\mathbb{K}}
\safemath{\opL}{\mathbb{L}}
\safemath{\opM}{\mathbb{M}}
\safemath{\opN}{\mathbb{N}}
\safemath{\opO}{\mathbb{O}}
\safemath{\opP}{\mathbb{P}}
\safemath{\opQ}{\mathbb{Q}}
\safemath{\opR}{\mathbb{R}}
\safemath{\opS}{\mathbb{S}}
\safemath{\opT}{\mathbb{T}}
\safemath{\opU}{\mathbb{U}}
\safemath{\opV}{\mathbb{V}}
\safemath{\opW}{\mathbb{W}}
\safemath{\opX}{\mathbb{X}}
\safemath{\opY}{\mathbb{Y}}
\safemath{\opZ}{\mathbb{Z}}
\safemath{\opZero}{\mathbb{O}}
\safemath{\identityop}{\opI}

\safemath{\veca}{\bma}
\safemath{\vecb}{\bmb}
\safemath{\vecc}{\bmc}
\safemath{\vecd}{\bmd}
\safemath{\vece}{\bme}
\safemath{\vecf}{\bmf}
\safemath{\vecg}{\bmg}
\safemath{\vech}{\bmh}
\safemath{\veci}{\bmi}
\safemath{\vecj}{\bmj}
\safemath{\veck}{\bmk}
\safemath{\vecl}{\bml}
\safemath{\vecm}{\bmm}
\safemath{\vecn}{\bmn}
\safemath{\veco}{\bmo}
\safemath{\vecp}{\bmp}
\safemath{\vecq}{\bmq}
\safemath{\vecr}{\bmr}
\safemath{\vecs}{\bms}
\safemath{\vect}{\bmt}
\safemath{\vecu}{\bmu}
\safemath{\vecv}{\bmv}
\safemath{\vecw}{\bmw}
\safemath{\vecx}{\bmx}
\safemath{\vecy}{\bmy}
\safemath{\vecz}{\bmz}

\safemath{\veczero}{\bmzero}
\safemath{\vecone}{\bmone}
\safemath{\vecxi}{\bmxi}
\safemath{\veclambda}{\bmlambda}
\safemath{\vecmu}{\bmmu}
\safemath{\vectheta}{\bmtheta}
\safemath{\vecphi}{\bmphi}
\safemath{\vecdelta}{\bmdelta}

\safemath{\matA}{\bA}
\safemath{\matB}{\bB}
\safemath{\matC}{\bC}
\safemath{\matD}{\bD}
\safemath{\matE}{\bE}
\safemath{\matF}{\bF}
\safemath{\matG}{\bG}
\safemath{\matH}{\bH}
\safemath{\matI}{\bI}
\safemath{\matJ}{\bJ}
\safemath{\matK}{\bK}
\safemath{\matL}{\bL}
\safemath{\matM}{\bM}
\safemath{\matN}{\bN}
\safemath{\matO}{\bO}
\safemath{\matP}{\bP}
\safemath{\matQ}{\bQ}
\safemath{\matR}{\bR}
\safemath{\matS}{\bS}
\safemath{\matT}{\bT}
\safemath{\matU}{\bU}
\safemath{\matV}{\bV}
\safemath{\matW}{\bW}
\safemath{\matX}{\bX}
\safemath{\matY}{\bY}
\safemath{\matZ}{\bZ}
\safemath{\matzero}{\bmzero}

\safemath{\matDelta}{\bDelta}
\safemath{\matLambda}{\bLambda}
\safemath{\matPhi}{\bPhi}
\safemath{\matSigma}{\bSigma}
\safemath{\matOmega}{\bOmega}
\safemath{\matTheta}{\bTheta}

\safemath{\matidentity}{\matI}
\safemath{\matone}{\matO}

\safemath{\rnda}{A}
\safemath{\rndb}{B}
\safemath{\rndc}{C}
\safemath{\rndd}{D}
\safemath{\rnde}{E}
\safemath{\rndf}{F}
\safemath{\rndg}{G}
\safemath{\rndh}{H}
\safemath{\rndi}{I}
\safemath{\rndj}{J}
\safemath{\rndk}{K}
\safemath{\rndl}{L}
\safemath{\rndm}{M}
\safemath{\rndn}{N}
\safemath{\rndo}{O}
\safemath{\rndp}{P}
\safemath{\rndq}{Q}
\safemath{\rndr}{R}
\safemath{\rnds}{S}
\safemath{\rndt}{T}
\safemath{\rndu}{U}
\safemath{\rndv}{V}
\safemath{\rndw}{W}
\safemath{\rndx}{X}
\safemath{\rndy}{Y}
\safemath{\rndz}{Z}

\safemath{\rveca}{\bimA}
\safemath{\rvecb}{\bimB}
\safemath{\rvecc}{\bimC}
\safemath{\rvecd}{\bimD}
\safemath{\rvece}{\bimE}
\safemath{\rvecf}{\bimF}
\safemath{\rvecg}{\bimG}
\safemath{\rvech}{\bimH}
\safemath{\rveci}{\bimI}
\safemath{\rvecj}{\bimJ}
\safemath{\rveck}{\bimK}
\safemath{\rvecl}{\bimL}
\safemath{\rvecm}{\bimM}
\safemath{\rvecn}{\bimN}
\safemath{\rveco}{\bomO}
\safemath{\rvecp}{\bimP}
\safemath{\rvecq}{\bimQ}
\safemath{\rvecr}{\bimR}
\safemath{\rvecs}{\bimS}
\safemath{\rvect}{\bimT}
\safemath{\rvecu}{\bimU}
\safemath{\rvecv}{\bimV}
\safemath{\rvecw}{\bimW}
\safemath{\rvecx}{\bimX}
\safemath{\rvecy}{\bimY}
\safemath{\rvecz}{\bimZ}

\safemath{\rvecxi}{\bmxi}
\safemath{\rveclambda}{\bmlambda}
\safemath{\rvecmu}{\bmmu}
\safemath{\rvectheta}{\bmtheta}
\safemath{\rvecphi}{\bmphi}

\safemath{\rmatA}{\bimA}
\safemath{\rmatB}{\bimB}
\safemath{\rmatC}{\bimC}
\safemath{\rmatD}{\bimD}
\safemath{\rmatE}{\bimE}
\safemath{\rmatF}{\bimF}
\safemath{\rmatG}{\bimG}
\safemath{\rmatH}{\bimH}
\safemath{\rmatI}{\bimI}
\safemath{\rmatJ}{\bimJ}
\safemath{\rmatK}{\bimK}
\safemath{\rmatL}{\bimL}
\safemath{\rmatM}{\bimM}
\safemath{\rmatN}{\bimN}
\safemath{\rmatO}{\bimO}
\safemath{\rmatP}{\bimP}
\safemath{\rmatQ}{\bimQ}
\safemath{\rmatR}{\bimR}
\safemath{\rmatS}{\bimS}
\safemath{\rmatT}{\bimT}
\safemath{\rmatU}{\bimU}
\safemath{\rmatV}{\bimV}
\safemath{\rmatW}{\bimW}
\safemath{\rmatX}{\bimX}
\safemath{\rmatY}{\bimY}
\safemath{\rmatZ}{\bimZ}

\safemath{\rmatDelta}{\bimDelta}
\safemath{\rmatLambda}{\bimLambda}
\safemath{\rmatPhi}{\bimPhi}
\safemath{\rmatSigma}{\bimSigma}
\safemath{\rmatOmega}{\bimOmega}
\safemath{\rmatTheta}{\bimTheta}

\usepackage{amssymb}
\usepackage{amsfonts}
\usepackage{mathrsfs}
\usepackage{xspace}
\usepackage{bm}
\usepackage{fancyref}
\usepackage{textcomp}

\usepackage{multirow}
\usepackage{stmaryrd}

\newenvironment{textbmatrix}{	\setlength{\arraycolsep}{2.5pt}%
								\big[\begin{matrix}}{\end{matrix}\big]%
								\raisebox{0.08ex}{\vphantom{M}}}

\def\be{\begin{equation}}
\def\ee{\end{equation}}
\def\een{\nonumber \end{equation}}
\def\mat{\begin{bmatrix}}
\def\emat{\end{bmatrix}}
\def\btm{\begin{textbmatrix}}
\def\etm{\end{textbmatrix}}

\def\ba#1\ea{\begin{align}#1\end{align}}
\def\bas#1\eas{\begin{align*}#1\end{align*}}
\def\bs#1\es{\begin{split}#1\end{split}}
\def\bg#1\eg{\begin{gather}#1\end{gather}}
\def\bml#1\eml{\begin{multline}#1\end{multline}}
\def\bi#1\ei{\begin{itemize}#1\end{itemize}}

\DeclareMathOperator*{\argmin}{arg\;min}		

\safemath{\dirac}{\delta}					
\safemath{\krond}{\dirac}					

\safemath{\upto}{\uparrow}
\safemath{\downto}{\downarrow}
\safemath{\iu}{j}							
\safemath{\ev}{\lambda}						
\safemath{\hilseqspace}{l^{2}}				
\newcommand{\banachfunspace}[1]{\setL^{#1}}	
\safemath{\hilfunspace}{\banachfunspace{2}}	

\safemath{\SNR}{\textit{SNR}} 				
\safemath{\PAR}{\textit{PAR}} 				
\safemath{\No}{N_0}							
\safemath{\Es}{E_s}							
\safemath{\Eb}{E_b}							
\safemath{\EbNo}{\frac{\Eb}{\No}}
\safemath{\EsNo}{\frac{\Es}{\No}}

\DeclareMathOperator{\CHop}{\ensuremath{\opH}} 
\safemath{\tvir}{\rndh_{\CHop}}				
\safemath{\tvtf}{\rndl_{\CHop}}				
\safemath{\spf}{\rnds_{\CHop}}				
\safemath{\bff}{H_{\CHop}}					

\safemath{\ircf}{r_{h}}						
\safemath{\tftvcf}{r_{s}}					
\safemath{\tfcf}{r_{l}}						
\safemath{\bfcf}{r_{H}}						

\safemath{\tcorr}{c_h}						
\safemath{\scf}{c_{s}}						
\safemath{\tfcorr}{c_{l}}					
\safemath{\fcorr}{c_{H}}						

\safemath{\mi}{I}							
\safemath{\capacity}{C}						

\safemath{\normal}{\mathcal{N}}			
\safemath{\jpg}{\mathcal{CN}}			
\safemath{\mchain}{\leftrightarrow}		

\safemath{\dB}{\,\mathrm{dB}}
\safemath{\dBm}{\,\mathrm{dBm}}
\safemath{\Hz}{\,\mathrm{Hz}}
\safemath{\kHz}{\,\mathrm{kHz}}
\safemath{\MHz}{\,\mathrm{MHz}}
\safemath{\GHz}{\,\mathrm{GHz}}
\safemath{\s}{\,\mathrm{s}}
\safemath{\ms}{\,\mathrm{ms}}
\safemath{\mus}{\,\mathrm{\text{\textmu}s}}
\safemath{\ns}{\,\mathrm{ns}}
\safemath{\ps}{\,\mathrm{ps}}
\safemath{\meter}{\,\mathrm{m}}
\safemath{\mm}{\,\mathrm{mm}}
\safemath{\cm}{\,\mathrm{cm}}
\safemath{\m}{\,\mathrm{m}}
\safemath{\W}{\,\mathrm{W}}
\safemath{\mW}{\, \mathrm{mW}}
\safemath{\J}{\,\mathrm{J}}
\safemath{\K}{\,\mathrm{K}}
\safemath{\bit}{\,\mathrm{bit}}
\safemath{\nat}{\,\mathrm{nat}}

\safemath{\define}{\triangleq}			

\safemath{\equivalent}{\sim}
\safemath{\distas}{\sim}					
\safemath{\sdiff}{\Delta}				

\safemath{\reals}{\mathbb{R}}
\safemath{\positivereals}{\reals_{+}}
\safemath{\integers}{\mathbb{Z}}
\safemath{\posint}{\integers_{+}}
\safemath{\naturals}{\mathbb{N}}
\safemath{\posnaturals}{\naturals_{+}}
\safemath{\complexset}{\mathbb{C}}
\safemath{\rationals}{\mathbb{Q}}

\newcommand*{\fancyrefapplabelprefix}{app}		
\newcommand*{\fancyrefthmlabelprefix}{thm}		
\newcommand*{\fancyreflemlabelprefix}{lem}		
\newcommand*{\fancyrefcorlabelprefix}{cor}		
\newcommand*{\fancyrefdeflabelprefix}{def}		
\newcommand*{\fancyrefproplabelprefix}{prop}		
\newcommand*{\fancyrefexmpllabelprefix}{exmpl}
\newcommand*{\fancyrefalglabelprefix}{alg}		
\newcommand*{\fancyreftbllabelprefix}{tbl}		

\frefformat{vario}{\fancyrefseclabelprefix}{Section~#1}
\frefformat{vario}{\fancyrefthmlabelprefix}{Theorem~#1}
\frefformat{vario}{\fancyreftbllabelprefix}{Table~#1}
\frefformat{vario}{\fancyreflemlabelprefix}{Lemma~#1}
\frefformat{vario}{\fancyrefcorlabelprefix}{Corollary~#1}
\frefformat{vario}{\fancyrefdeflabelprefix}{Definition~#1}
\frefformat{vario}{\fancyreffiglabelprefix}{Fig.~#1}
\frefformat{vario}{\fancyrefapplabelprefix}{Appendix~#1}
\frefformat{vario}{\fancyrefeqlabelprefix}{(#1)}
\frefformat{vario}{\fancyrefproplabelprefix}{Proposition~#1}
\frefformat{vario}{\fancyrefexmpllabelprefix}{Example~#1}
\frefformat{vario}{\fancyrefalglabelprefix}{Algorithm~#1}

\safemath{\dictab}{[\,\dicta\,\,\dictb\,]}

\safemath{\ysig}{\bmy}
\safemath{\ysighat}{\hat{\ysig}}
\safemath{\ysigdim}{M}
\safemath{\xsig}{\bmx}
\safemath{\xsigdim}{N}
\safemath{\nx}{n_x}
\safemath{\zsig}{\bmz}
\safemath{\zsigdim}{\ysigdim}
\safemath{\rsig}{\bmr}
\safemath{\Adict}{\bA}
\safemath{\Adicttilde}{\widetilde{\Adict}}
\safemath{\Adictdim}{\outputdim\times\xsigdim}
\safemath{\avec}{\bma}
\safemath{\avectilde}{\tilde{\avec}}
\safemath{\Bdict}{\bB}
\safemath{\Bdicttilde}{\widetilde{\Bdict}}
\safemath{\Cdict}{\bC}
\safemath{\cvec}{\bmc}
\safemath{\Ddict}{\bD}
\safemath{\Ddictdim}{\ysigdim\times\xsigdim}
\safemath{\dvec}{\bmd}
\safemath{\Ddicttilde}{\widetilde{\bD}}
\safemath{\Bonb}{\bB}
\safemath{\bvec}{\bmb}
\safemath{\Bonbdim}{\ysigdim\times\ysigdim}
\safemath{\noise}{\bmn}
\safemath{\noisedim}{\ysigim}
\safemath{\err}{\bme}
\safemath{\errdim}{\ysigdim}
\safemath{\errset}{\setE}
\safemath{\nerr}{n_e}
\safemath{\delop}{\bP_\errset}
\safemath{\delopc}{\bP_{{\errset}^c}}

\safemath{\cplxi}{\imath}
\safemath{\cplxj}{\jmath}

\safemath{\dict}{\matD}
\safemath{\inputdim}{N}		
\safemath{\outputdim}{M}		
\safemath{\sparsity}{S}	
\safemath{\inputdimA}{{N_a}}	
\safemath{\inputdimB}{{N_b}}	
\safemath{\elemA}{{n_a}}	
\safemath{\elemB}{{n_b}}	
\safemath{\resA}{\matR_a}	
\safemath{\resB}{\matR_b}	
\safemath{\subD}{\matS} 
\safemath{\subA}{\matS_a} 
\safemath{\subB}{\matS_b} 
\safemath{\dicta}{\matA} 	
\safemath{\dictb}{\matB} 	
\safemath{\hollowS}{H}
\safemath{\hollowA}{H_a}
\safemath{\hollowB}{H_b}
\safemath{\cross}{Z}
\safemath{\coh}{\mu_d}			
\safemath{\coha}{\mu_a}			
\safemath{\cohb}{\mu_b}			
\safemath{\mubs}{\nu}	
\safemath{\cohm}{\mu_m} 
\safemath{\dictset}{\setD}	
\safemath{\dictsetp}{\dictset(\coh,\coha,\cohb)}	
\safemath{\dictsetgen}{\dictset_\text{gen}}
\safemath{\dictsetgenp}{\dictsetgen(\coh)}
\safemath{\dictsetonb}{\dictset_\text{onb}}
\safemath{\dictsetonbp}{\dictsetonb(\coh)}

\safemath{\leftside}{U}
\safemath{\rightsideA}{R_a}
\safemath{\rightsideB}{R_b}

\safemath{\indexS}{\setI_S} 

\safemath{\na}{n_a}			
\safemath{\nb}{n_b}			
\safemath{\coeffa}{p_i}	
\safemath{\coeffb}{q_j}	
\safemath{\seta}{\setP}		
\safemath{\setb}{\setQ}     
\safemath{\setw}{\setW}	
\safemath{\setz}{\setZ}	
\safemath{\cola}{\veca}		
\safemath{\colb}{\vecb}		
\safemath{\cold}{\vecd}		
\safemath{\inputvec}{\vecx} 	
\safemath{\error}{\vece}	
\safemath{\noiseout}{\vecz} 	
\safemath{\inputvecel}{x}
\safemath{\inputveca}{\vecx_a}
\safemath{\inputvecb}{\vecx_b}
\safemath{\outputvec}{\vecy}	
\safemath{\lambdamin}{\lambda_{\mathrm{min}}}

\safemath{\elltwo}{\ell_2}
\safemath{\ellone}{\ell_1}
\safemath{\ellzero}{\ell_0}
\safemath{\ellinf}{\ell_\infty}
\safemath{\ellinftilde}{\ell_{\widetilde\infty}}
\safemath{\licard}{Z(\coh,\coha,\cohb)}
\safemath{\xsol}{\hat{x}}
\safemath{\xbord}{x_b}		
\safemath{\xstat}{x_s}		
\safemath{\xstatLone}{\tilde{x}_s}
\safemath{\order}{\mathcal{O}} 
\safemath{\scales}{\Theta} 
\safemath{\ones}{\mathbf{1}} 
\safemath{\zeroes}{\mathbf{0}} 
\safemath{\thlone}{\kappa(\coh,\cohb)} 
\safemath{\constoneA}{\delta} 
\safemath{\constoneB}{\epsilon} 
\safemath{\nlarge}{L}				   
\safemath{\sumlarge}{S_\nlarge}
\safemath{\maxlarger}{P_\nlarge}	   
\safemath{\Pzero}{\textrm{P0}}	
\safemath{\Pone}{\textrm{P1}}
\safemath{\vecfir}{\vecw}			 
\safemath{\vecsec}{\vecz}
\safemath{\elvecfir}{w}              
\safemath{\elvecsec}{z}				 
\safemath{\nlargefir}{n}
\safemath{\normout}{\gamma}
\safemath{\auxfun}{h}
\safemath{\supp}{\textrm{supp}}

\safemath{\indexa}{\ell}
\safemath{\indexb}{r}
\safemath{\indexc}{i}
\safemath{\indexd}{j}

\safemath{\project}{P}

\begin{document}

\title{Joint Active User Detection, Channel Estimation, and Data Detection for Massive Grant-Free Transmission in Cell-Free Systems}

\author{\IEEEauthorblockN{Gangle~Sun\textsuperscript{1,2}, Mengyao~Cao\textsuperscript{1}, Wenjin~Wang\textsuperscript{1,2}, Wei~Xu\textsuperscript{1,2}, and Christoph~Studer\textsuperscript{3}}\\
	\textit{\textsuperscript{1}National Mobile Communications Research Laboratory, Southeast University, Nanjing, China}\\
	\textit{\textsuperscript{2}Purple Mountain Laboratories, Nanjing, China}\\
	\textit{\textsuperscript{3}Department of Information Technology and Electrical Engineering, ETH Zurich, Switzerland}\\
	\textit{email: sungangle@seu.edu.cn, cmengyao64@gmail.com, \{wangwj,\,wxu\}@seu.edu.cn, and studer@ethz.ch}\\
\thanks{This work was supported in part by the National Natural Science Foundation of China (NSFC) under Grants 62341110, 62022026 and 62211530108; in part by the Jiangsu Province Basic Research Project under Grant BK20192002; in part by the Fundamental Research Funds for the Central Universities under Grants 2242022k30005, 2242022k60002 and 2242023k5003.  The work of Gangle Sun was supported in part
	by China Scholarship Council (CSC) under Grant 202206090074. The work of CS was supported in part by the U.S. National Science Foundation (NSF) 
under grants CNS-1717559 and ECCS-1824379, and in part by an ETH Research Grant.}
	\thanks{The authors would like to thank Victoria Palhares and Haochuan Song for discussing channel modeling of cell-free systems. We also acknowledge Gian Marti and Sueda Taner for their suggestions on deriving the FBS algorithm.}
}

\maketitle

\begin{abstract}
Cell-free communication has the potential to significantly improve grant-free transmission in massive machine-type communication, wherein multiple access points jointly serve a large number of user equipments to improve coverage and spectral efficiency. In this paper, we propose a novel framework for joint active user detection (AUD), channel estimation (CE), and data detection (DD) for massive grant-free transmission in cell-free systems. We formulate an optimization problem for joint AUD, CE, and DD by considering both the sparsity of the data matrix, which arises from intermittent user activity, and the sparsity of the effective channel matrix, which arises from intermittent user activity and large-scale fading. We approximately solve this optimization problem with a box-constrained forward-backward splitting algorithm, which significantly improves AUD, CE, and DD performance. We demonstrate the effectiveness of the proposed framework through simulation experiments.
\end{abstract}
\section{Introduction}
Massive machine-type communications (mMTC) is a central scenario in fifth-generation (5G) wireless communication systems, in which user equipments (UEs) transmit data intermittently to an infrastructure base station (BS).
Massive grant-free transmission techniques are suitable for mMTC scenarios as they reduce excessive signaling overhead, network congestion, and high transmission latency by allowing the active UEs to transmit signals over shared resource elements directly without sophisticated scheduling mechanisms~\cite{Sun2022massive}.

In order to improve coverage for UEs in mMTC scenarios, cell-free communication techniques have emerged as a powerful solution \cite{Mishra2022rate,xu2023toward}.
Cell-free communication mitigates inter-cell interference and improves spectral efficiency by jointly processing all of the information acquired at a large number of distributed access points (APs) that are connected to a central processing unit (CPU)~\cite{song2022joint,Ke2021massive,xu2023edge}.
One of the key tasks for massive grant-free transmission in cell-free systems involves detecting the set of active UEs, estimating their channels, and detecting their transmitted data at the CPU side.

\subsection{Contributions}
This paper proposes a novel framework for joint active user detection, channel estimation, and data detection (JACD) suitable for massive grant-free transmission in cell-free systems. 
We formulate the JACD problem as a nonconvex optimization problem, accounting for sparsity in the data matrix arising from UEs' sporadic activity and sparsity in the effective channel matrix resulting from both the UEs' sporadic activity and  large-scale fading.
We relax the discrete constellation constraints in our problem formulation, which enables the use of computationally-efficient gradient-type solvers.
We then develop a forward-backward splitting (FBS) algorithm to approximately solve the JACD problem, leading to a significant improvement in joint estimation accuracy.
Finally, we demonstrate the effectiveness of the proposed algorithm in terms of active user detection (AUD), channel estimation (CE), and data detection (DD) through system simulations.

\subsection{Relevant Prior Art}

Recent research has focused on  AUD, CE, and DD for massive grant-free transmission in cell-free communication systems \cite{Ganesan2021algorithm, Jiang2023EM, guo2022joint, Ke2021massive, wang2022two, Iimori2021grant}.
Reference~\cite{Ganesan2021algorithm} proposed two different AUD algorithms based on dominant APs and clustering, respectively, demonstrating that cell-free communication can surpass co-located schemes for AUD in large coverage areas.
Reference~\cite{Jiang2023EM} proposed an expectation-maximization approximate message passing (AMP) algorithm for CE and detected active UEs using posterior support probabilities.
Reference \cite{guo2022joint} developed a Bayesian AMP algorithm based on a single measurement vector for joint AUD and CE, processing received signals at each AP separately.
Reference \cite{Ke2021massive} presented an  AMP algorithm for joint AUD and CE, accounting for quantization artifacts and exploited the sparsity structure in the channel matrix.
Reference \cite{wang2022two} introduced a two-stage CE and AUD method using the vector AMP algorithm for AUD followed by linear minimum mean square error-based CE.
Reference~\cite{Iimori2021grant} introduced a bilinear Gaussian belief propagation algorithm for JACD, combining successive interference cancellation and Bayesian message passing methods. 
Unlike most prior works, we address the JACD problems for mMTC in cell-free systems and exploit sparsity in the data matrix arising from intermittent UE activity to improve the joint estimation performance.

JACD for single-cell massive grant-free transmission has been explored as well in~\cite{zou2020alow, Di2022joint, zhang2023joint, jiang2020joint}.
Reference \cite{zou2020alow} applied bilinear generalized AMP (BiG-AMP) and belief propagation algorithms for JACD in massive grant-free systems with low-precision data converters.
Reference  \cite{Di2022joint} proposed the use of bilinear message-scheduling generalized AMP to enable JACD and data detection by utilizing channel decoder beliefs to improve AUD and detection performance.
Reference \cite{zhang2023joint} developed a BiG-AMP algorithm based on the row-sparse channel matrix structure for JACD by leveraging channel correlation across different antennas.
Reference \cite{jiang2020joint} utilized AMP to decouple transmissions of different UEs and addressed the nonlinear coupling of each UE's activity, channel coefficient, and data separately. 
In contrast, our work focuses on JACD in cell-free systems, utilizes an optimization-based problem formulation, deploys gradient-type algorithms, and considers both sources of sparsity in the channel matrix due to sporadic UE activity in mMTC scenarios and large-scale fading in cell-free wireless communication systems.

\subsection{Notation}
Uppercase and lowercase boldface letters denote matrices and column vectors, respectively;
$\mathbf{A}(m,n)$ and $\mathbf{a}(m)$ correspond to the element in the $m$th row and $n$th column of the matrix~$\mathbf{A}$ and the $m$th
element of the vector $\mathbf{a}$, respectively;
$\mathbf{1}_{M\times N}$ represents the $M\times N$ all-ones matrix.
The superscripts $\!\left(\cdot\right)\!^T$ and $\!\left(\cdot\right)\!^H$ denote transpose and conjugate transpose, respectively; $\|\cdot\|_F$ is the Frobenius norm, $|\mathcal{Q}|$ the number of elements in the set $\mathcal{Q}$, and $\odot$ the Hadamard product.
Proportional relationships are denoted by $\propto$.
The indicator function $\mathbb{I}\left\{\cdot\right\}$ is $1$ if the condition is true and $0$ otherwise; 
$\text{Re}\{x\}$ and $\text{Im}\{x\}$ are the real and image parts of~$x\in\complexset$, respectively. $\mathbb{P}\{\cdot\}$ denotes probability. 

\section{System Model}
We consider a cell-free mMTC system consisting of $P$ distributed APs with $M$ antennas each and $N$ single-antenna UEs, where
$N_a \ll N$ active UEs transmit signals simultaneously to the APs over  $R$ shared resource elements. 
In what follows, we assume that all UEs are perfectly synchronized.

We consider frequency-flat and block-fading channels with the following input-output relation \cite{sun2021OFDMA,Ke2021massive,Sun2022massive}:
\begin{equation}
	\textstyle
		\mathbf{Y} = \sum_{n=1}^{N}\xi_n \mathbf{h}_{n}\mathbf{x}_n^T + \mathbf{N}.
	\label{1}
\end{equation}
Here, $\mathbf{Y}\in\mathbb{C}^{MP\times R}$ contains the received signals of all APs for the $R$ resource elements, $\xi_n\in\{0,1\}$ is $n$th UE's activity indicator with $\xi_n=1$ if the $n$th UE is active and $\xi_n=0$ otherwise. 
The vector $\mathbf{h}_{n}=[\mathbf{h}_{n,1}^T,\mathbf{h}_{n,2}^T,\ldots,\mathbf{h}_{n,P}^T]^T\in\mathbb{C}^{MP\times 1}$ represents the channel vector between the $n$th UE and all APs while $\mathbf{h}_{n,p}\in \mathbb{C}^{M\times 1}$ is the channel vector between the $n$th UE and the $p$th AP, following the model in  \cite{song2022joint,ngo2017cell}.  
The vector $\mathbf{x}_n = \left[\mathbf{x}_{\text{P},n}^T,\mathbf{x}_{\text{D},n}^T\right]^T\in\mathbb{C}^{R\times 1}$ contains the $n$th UE's pilots $\mathbf{x}_{\text{P},n}\in\mathbb{C}^{R_P\times 1}$ and data signals $\mathbf{x}_{\text{D},n}\in\mathcal{Q}^{R_D\times 1}$, where $R = R_P+R_D$ and $\mathcal{Q}$ is a discrete constellation set. 
The matrix $\mathbf{N}\in\mathbb{C}^{MP\times R}$ models noise with i.i.d. circularly-symmetric complex Gaussian entries of variance $\No=1$.

\begin{figure}[tp]
	\centering
	\includegraphics[width=0.33\textwidth]{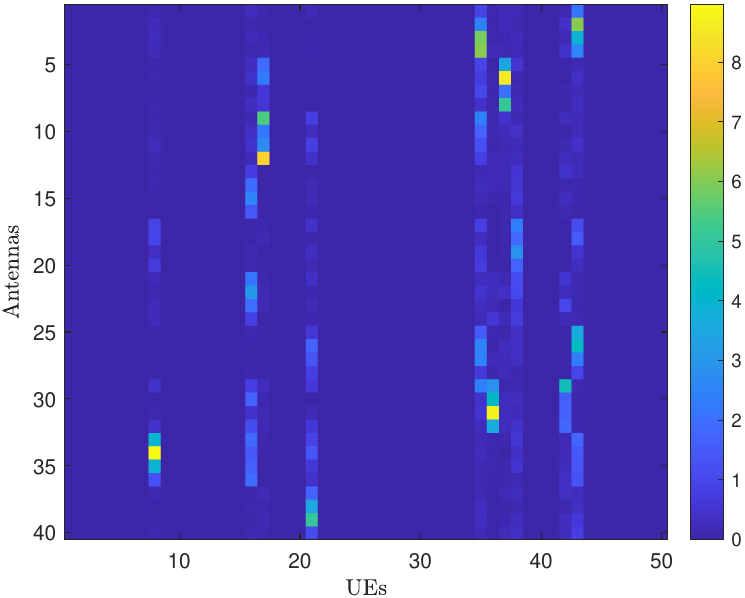}
	\caption{Amplitude of entries of the efficient channel matrix $\mathbf{H}$ between $10$ APs with $4$ antennas each and $50$ UEs, where only $10$ UEs are active.}
	\label{location}
\end{figure}

Since $\xi_n\mathbf{h}_n\mathbf{x}_{\text{D},n}^T=(\xi_n\mathbf{h}_n)(\xi_n\mathbf{x}_{\text{D},n}^T)$, the data vector $\mathbf{x}_{\text{D},n}$ can also be treated as a zero vector if $n$th UE is inactive. 
As such, we can rewrite (\ref{1}) as follows: 
\begin{align}
	\textstyle
	\mathbf{Y} = \mathbf{H}\left[\mathbf{X}_{\text{P}},\mathbf{X}_{\text{D}}\right]+\mathbf{N}= \mathbf{H}\mathbf{X}+\mathbf{N}.
	\label{pilot_data}
\end{align}
Here, $\mathbf{H}\triangleq\left[\xi_1\mathbf{h}_{1},\xi_2\mathbf{h}_{2},\ldots,\xi_N\mathbf{h}_{N}\right]\in\mathbb{C}^{MP\times N}$ is the effective channel matrix, $\mathbf{X}_{\text{P}} = [\mathbf{x}_{\text{P},1},\mathbf{x}_{\text{P},2},\ldots,\mathbf{x}_{\text{P},N}]^T\in\mathbb{C}^{N\times R_P}$ is the pilot matrix, and $\mathbf{X}_{\text{D}} = [\bar{\mathbf{x}}_{\text{D},1},\bar{\mathbf{x}}_{\text{D},2},\ldots,\bar{\mathbf{x}}_{\text{D},N}]^T\in\bar{\mathcal{Q}}^{N\times R_D}$ is the sparse data matrix, where $\bar{\mathbf{x}}_{\text{D},n}\triangleq\xi_n\mathbf{x}_{\text{D},n}$ and $\bar{\mathcal{Q}}\triangleq\left\{\mathcal{Q},0\right\}$. To simplify notation, we define $\mathbf{X}\triangleq\left[\mathbf{X}_{\text{P}},\mathbf{X}_{\text{D}}\right]$ in (\ref{pilot_data}). 
Note that, in signal matrix $\mathbf{X}$, we only consider sparsity in the data matrix $\mathbf{X}_{\text{D}}$ and not in the pilot matrix $\mathbf{X}_{\text{P}}$ as our optimization problem will leverage all of the available pilot information---nonetheless, inactive UEs will not transmit any pilots.
In addition, as illustrated in Fig.~\ref{location}, there are two sources of sparsity in the effective channel matrix~$\mathbf{H}$: (i) column sparsity caused by the UEs' sporadic activity and (ii) inherent channel sparsity among different APs caused by the fact that each UE is only nearby a few APs. In the following, we will explore both of these sources of sparsity.

\section{Joint Active User Detection, Channel Estimation, and Data Detection}
\newcounter{TempEqCnt}
\setcounter{TempEqCnt}{\value{equation}}
\setcounter{equation}{5}
\begin{figure*}
	\vspace{-1ex}
	\begin{equation}
		\begin{aligned}
			\textstyle\mathcal{P}_1:\;\left\{ {\hat{\mathbf{H}},{{\hat {\mathbf{X}}}_{\text{D}}}} \right\} =\mathop {\argmin }\textstyle_{\scriptstyle{\mathbf{H}\in\mathbb{C}^{MP\times N}}\hfill\atop
				{	\scriptstyle{{\mathbf{X}}_{\text{D}}\in\bar{\mathcal{Q}}^{N\times R_D}}\hfill}} \frac{1}{2} \left\|\mathbf{Y}-\mathbf{H}\left[\mathbf{X}_{\text{P}},\mathbf{X}_{\text{D}}\right]\right\|_F^2 
			+ \mu_h\sum_{n=1}^N\sum_{p=1}^P\left\|\mathbf{h}_{n,p}\right\|_F+ \mu_x\sum_{n=1}^N\left\|\bar{\mathbf{x}}_{\text{D},n}\right\|_F.
		\end{aligned}
		\label{P1}
	\end{equation}
	\vspace{-3ex}
	\hrulefill
\end{figure*}
\setcounter{equation}{\value{TempEqCnt}}

\subsection{Problem Formulation}
According to system model (\ref{pilot_data}), the JACD optimization problem for mMTC in cell-free systems can be formulated as
\begin{align}
	\textstyle
\Big\{ \hat{\mathbf{H}},{{\hat {\mathbf{X}}}_{\text{D}}} \Big\} =\mathop {\arg \max }\limits_{\scriptstyle{\mathbf{H}\in\mathbb{C}^{MP\times N}}\hfill\atop
			{	\scriptstyle{{\mathbf{X}}_{\text{D}}\in\bar{\mathcal{Q}}^{N\times R_D}}\hfill}} P\!\left(\mathbf{Y}|\mathbf{H},\mathbf{X}_{\text{D}}\right)\!P\!\left(\mathbf{H}\right)\!P\!\left(\mathbf{X}_{\text{D}}\right)
	\label{joint_problem}
\end{align}
with the channel law
\begin{align}
	\textstyle P\!\left(\mathbf{Y}|\mathbf{H},\mathbf{X}_{\text{D}}\right) \propto\exp\!\left(-\left\|\mathbf{Y}-\mathbf{H}\left[\mathbf{X}_{\text{P}},\mathbf{X}_{\text{D}}\right]\right\|_F^2/\No\right)\!.
	\label{P_Y_given_H_X_D}
\end{align}
To take into account the sparsity in both $\mathbf{H}$ and $\mathbf{X}_{\text{D}}$, we utilize the following  models for $\mathbf{H}$ and $\mathbf{X}_{\text{D}}$:
\begin{subequations}
	\label{P_H_P_X_D}
	\begin{align}
		&\textstyle
		P\!\left(\mathbf{H}\right) \propto \prod_{n=1}^{N}\prod_{p=1}^{P}\exp\!\left(-2\mu_h\left\|\mathbf{h}_{n,p}\right\|_F\right)\!,\label{P_H}\\
		&\textstyle
		P\!\left(\mathbf{X}_{\text{D}}\right) \propto \prod_{n=1}^{N}\exp\!\left(-2\mu_x\left\|\bar{\mathbf{x}}_{\text{D},n}\right\|_F\right)\!, \mathbf{X}_{\text{D}}\in\bar{\mathcal{Q}}^{N\times R_D},\label{P_D}
	\end{align}
\end{subequations}
where $\mu_h$ and $\mu_x$ are parameters that determine the amount of sparsity. 
The sparsity in $\mathbf{H}$ due to UE activity and large-scale fading is modeled by a complex-valued block-Laplace prior in (\ref{P_H}), whereas the sparsity of $\mathbf{X}_{\text{D}}$ due to UE activity is modeled by a complex-valued Laplace prior in (\ref{P_D}) \cite{song2022joint}.
Finally, by plugging (\ref{P_Y_given_H_X_D}) and (\ref{P_H_P_X_D}) into (\ref{joint_problem}) and taking the logarithm, we obtain the optimization problem $\mathcal{P}_1$ in (\ref{P1}).

\subsection{Problem Relaxation}
The discrete set $\bar{\mathcal{Q}}^{N\times R_D}$ of $\mathbf{X}_{\text{D}}$ renders $\mathcal{P}_1$ a discrete-valued optimization problem, and a na\"ive exhaustive search would be impractical. 
To arrive at a tractable optimization problem, we relax the set $\bar{\mathcal{Q}}$ to its convex hull $\mathcal{B}$ as in \cite{song2022joint}
\setcounter{equation}{6}
\begin{equation}
	\textstyle
	\mathcal{B}=\left\{\sum_{i=1}^{|\bar{\mathcal{Q}|}}\delta_i q_i:\,q_i\in\bar{\mathcal{Q}},\,\delta_i\ge 0,\forall i;\,\sum_{i=1}^{|\bar{\mathcal{Q}|}}\delta_i=1\right\}\!.
\end{equation}
For simplicity of exposition, we consider quadrature phase shift keying (QPSK) with $\mathcal{B}=\!\{x\in\mathbb{C}:\,-B\le\text{Re}\{x\}\le B,\,-B\le\text{Im}\{x\}\le B\}\!$ and $B>0$. Nonetheless, our method can be generalized to other, higher-order quadrature amplitude modulation constellation sets. 

By relaxing $\mathcal{P}_1$ into a continuous optimization problem, we still want the estimated elements of $\mathbf{X}_{\text{D}}$ to be in the set~$\bar{\mathcal{Q}}$. 
To this end, we introduce a regularizer $\mathcal{C}\!\left(\mathbf{X}_{\text{D}}\right)\!=-\|\mathbf{X}_{\text{D}}\odot\mathbf{X}_{\text{D}}^*-B^2\mathbf{1}_{N\times R_D}\|_F^2$ to $\mathcal{P}_1$, which pushes elements of $\mathbf{X}_{\text{D}}$ into $\bar{\mathcal{Q}}$~\cite{Oscar2017bit}, where $\mathcal{C}\!\left(\mathbf{X}_{\text{D}}\right)\!$ will take its minimum only if all elements in~$\mathbf{X}_{\text{D}}$ fall in $\bar{\mathcal{Q}}$. As such, we can transform $\mathcal{P}_1$ into 
\begin{equation}
	\begin{aligned}
		\textstyle\mathcal{P}_2:\;\left\{ {\hat{\mathbf{H}},{{\hat {\mathbf{X}}}_{\text{D}}}} \right\} =&\mathop {\argmin}_{\scriptstyle{\mathbf{H}\in\mathbb{C}^{MP\times N}}\hfill\atop
			{	\scriptstyle{{\mathbf{X}}_{\text{D}}\in{\mathcal{B}}^{N\times R_D}}\hfill}} \textstyle\frac{1}{2}\left\|\mathbf{Y}-\mathbf{H}\left[\mathbf{X}_{\text{P}},\mathbf{X}_{\text{D}}\right]\right\|_F^2 \\ 
		&\quad\textstyle+ \mu_h\sum_{n=1}^N\sum_{p=1}^P\|\mathbf{h}_{n,p}\|_F\\
		&\quad\textstyle
		+ \mu_x\sum_{n=1}^N\|\bar{\mathbf{x}}_{\text{D},n}\|_F+\lambda\,\mathcal{C}\!\left(\mathbf{X}_{\text{D}}\right)\!,
	\end{aligned}
	\label{P2}
\end{equation}
which can be solved approximately with FBS~\cite{goldstein2014field,beck2009fast}.

\subsection{Box-Constrained FBS Algorithm}

FBS splits the objective function of a convex optimization problem into a smooth function $f(\mathbf{S})$ and an arbitrary (not necessarily smooth) function $g(\mathbf{S})$\mbox{\cite{goldstein2014field,beck2009fast}}:
\begin{equation}
	\begin{aligned}
		\hat{\mathbf{S}} = \argmin\textstyle_{\mathbf{S}}  \;f(\mathbf{S})
		+ g(\mathbf{S}).
	\end{aligned}
\end{equation}
The principle is to alternate between a gradient step for the smooth function $f(\mathbf{S})$ ({forward step}) and a proximal operation to find a point near the minimizer of the non-smooth function $g(\mathbf{S})$ ({backward step}).
This process is repeated until convergence (e.g., a stopping criterion is met). We apply this technique to the nonconvex problem $\mathcal{P}_2$. 

\subsubsection{Problem Splitting}
With the definition $\mathbf{S}\triangleq[\mathbf{H}^H, \mathbf{X}_{\text{D}}]^H\in\mathbb{C}^{(MP+R_D)\times N}$, we split the objective function in $\mathcal{P}_2$ into
\begin{equation}
	\begin{aligned}
		f(\mathbf{S})=&\textstyle\frac{1}{2}\|\mathbf{Y}-\mathbf{H}[\mathbf{X}_{\text{P}},\mathbf{X}_{\text{D}}]\|_F^2  + \lambda\mathcal{C}(\mathbf{X}_{\text{D}}),\\
		g(\mathbf{S})=&\textstyle\mu_h\sum_{n=1}^N\sum_{p=1}^P\|\mathbf{h}_{n,p}\|_F + \mu_x\sum_{n=1}^N\|\bar{\mathbf{x}}_{\text{D},n}\|_F\\
		&+ \mathcal{X}(\mathbf{X}_{\text{D}}),
	\end{aligned}
\end{equation}
where $\mathcal{X}\!\left(\mathbf{X}_{\text{D}}\right)\!$ enforces the data to be within the convex set $\mathcal{B}$:
\begin{equation}
	\mathcal{X}\!\left(\mathbf{X}_{\text{D}}\right)\!\triangleq\left\{\begin{array}{cl}
		+\infty &, \; \exists \mathbf{X}_{\text{D}}(n,r) \notin \mathcal{B},\;\forall n,r \\
		0 &, \; \mathbf{X}_{\text{D}}(n,r) \in \mathcal{B},\;\forall n,r.
	\end{array}\right.
\end{equation}

\subsubsection{Forward Step}
The forward step is given by
\begin{equation}
	\hat{\mathbf{S}}^{k}=\mathbf{S}^k-\tau^k \nabla f \!\left(\mathbf{S}^k\right)\!,
\end{equation}
where the superscript $k$ indicates the $k$th iteration, $\hat{\mathbf{S}}^{k} = [(\hat{\mathbf{H}}^{k})^H,\hat{\mathbf{X}}_{\text{D}}^{k}]^H$, $\tau^k$ is the step size of the $k$th iteration~\cite{goldstein2014field}, and gradient of $f \!\left(\mathbf{S}\right)\!$ with respect to $\mathbf{S}$ is given by $\nabla f (\mathbf{S}) =[(\frac{\partial f }{\partial \mathbf{H}^*} )^T,(\frac{\partial f }{\partial \mathbf{X}_{\text{D}}^T} )^T]^T$ with
\begin{equation}
	\begin{aligned}
		\textstyle \frac{\partial f }{\partial \mathbf{H}^*}=&-\!\left(\mathbf{Y}-\mathbf{H}\mathbf{X}\right)\!\mathbf{X}^H,\\
		\textstyle\frac{\partial f }{\partial \mathbf{X}_{\text{D}}^T}=&\textstyle-\!\left(\mathbf{Y}_{\text{D}}-\mathbf{H}\mathbf{X}_{\text{D}}\right)\!^H\mathbf{H} + \lambda\frac{\partial \mathcal{C}\!\left(\mathbf{X}_{\text{D}}\right)\!}{\partial\mathbf{X}_{\text{D}}^T},
	\end{aligned}
\end{equation}
where $
\frac{\partial \mathcal{C}\!\left(\mathbf{X}_{\text{D}}\right)\!}{\partial \mathbf{X}_{\text{D}}^T}=-4 (\mathbf{X}_{\text{D}}^*\odot(\mathbf{X}_{\text{D}}\odot\mathbf{X}_{\text{D}}^*-B^2\mathbf{1}_{N\times R_D}))^T.$ 

\subsubsection{Backward Step}
The proximal operator for $\mathbf{H}$ is  
\begin{equation}
	\begin{aligned}
		\mathbf{H}^{k+1}=\argmin\textstyle_{\mathbf{H}\in\mathbb{C}^{MP\times N}}  &\textstyle\;\frac{1}{2} \|\mathbf{H}-\hat{\mathbf{H}}^{k}\|_F^2 \\
		+ \tau^k\mu_h&\textstyle\sum_{n=1}^N\sum_{p=1}^P\left\|\mathbf{h}_{n,p}\right\|_F,
	\end{aligned}
\end{equation}
which has the following closed-form solution \cite{song2022joint,goldstein2014field,beck2009fast}:
\begin{equation}
	\mathbf{h}_{n,p}^{k+1} = \hat{\mathbf{h}}_{n,p}^{k}\textstyle\frac{\max\{\|\hat{\mathbf{h}}_{n,p}^{k}\|_F-\tau^k\mu_h,0\}}{\|\hat{\mathbf{h}}_{n,p}^{k}\|_F}.
	\label{Shrinkage}
\end{equation}
The proximal operator for $\mathbf{X}_{\text{D}}$ is given by
\begin{equation}
	\begin{aligned}
		\mathbf{X}_{\text{D}}^{k+1}=\argmin\textstyle_{\mathbf{X}_{\text{D}}\in\mathcal{B}^{N\times R_D}} &\textstyle\;\frac{1}{2} \|\mathbf{X}_{\text{D}}-\hat{\mathbf{X}}_{\text{D}}^{k}\|_F^2\\
		+  \tau^k\mu_x&\textstyle\sum_{n=1}^N\left\|\bar{\mathbf{x}}_{\text{D},n}\right\|_F,
	\end{aligned}
\label{Xd_prox}
\end{equation}
which is a convex optimization problem.
To obtain the optimal solution of (\ref{Xd_prox}), we first decompose the problem (\ref{Xd_prox}) as $N$ independent subproblems
\begin{equation}
	\begin{aligned}
		&\textstyle\mathbf{r}_{n} = \argmin_{\mathbf{r}_{n}\in\mathbb{C}^{2R_D\times 1}}   \frac{1}{2} \|\mathbf{r}_{n}-\hat{\mathbf{r}}_{n}^{k}\|_F^2 + \tau^k\mu_x\|\mathbf{r}_{n}\|_F\\
		&\textstyle\text{s.t. } -B \le \mathbf{r}_{n}\!\left(d\right)\!\le B,\;\forall d\in\left\{1,2,3,\ldots,2R_D\right\},
	\end{aligned}
\label{r_n}
\end{equation}
where $\mathbf{r}_n \triangleq [\text{Re}\left\{\bar{\mathbf{x}}_{\text{D},n}\right\}^T,\text{Im}\left\{\bar{\mathbf{x}}_{\text{D},n}\right\}^T]^T\in\mathbb{R}^{2R_D\times 1}$ and $\hat{\mathbf{r}}_n^{k} \triangleq [\text{Re}\{\hat{\mathbf{x}}_{\text{D},n}^{k}\}^T,\text{Im}\{\hat{\mathbf{x}}_{\text{D},n}^{k}\}^T]^T\in\mathbb{R}^{2R_D\times 1}$. 
Let $L(\mathbf{r}_n,\mathbf{p},\mathbf{q})=\frac{1}{2} \|\mathbf{r}_{n}-\hat{\mathbf{r}}_{n}^{k}\|_F^2 +\tau^k\mu_x\|\mathbf{r}_{n}\|_F +  \sum_{d}\mathbf{p}(d)(\mathbf{r}_{n}(d)-B)- \sum_{d}\mathbf{q}(d)(\mathbf{r}_{n}(d)+B)$ be the Lagrangian, then the Karush-Kuhn-Tucker (KKT) conditions are as follows:
\begin{subequations}
	\begin{align}
		& \textstyle\mathbf{r}_n + \frac{\tau^k \mu_x}{\left\|\mathbf{r}_n\right\|_F}\mathbf{r}_n-\hat{\mathbf{r}}_n^{k}+\mathbf{p}-\mathbf{q}=0,\label{40a}\\
		&\mathbf{r}_n(d)-B \le 0,\;-\mathbf{r}_n(d)-B \le 0,\;\forall d,\\
		&\mathbf{p}(d)\ge0,\;\mathbf{q}(d)\ge0,\;\forall d,\\
		&\mathbf{p}(d)(\mathbf{r}_n(d)-B)=0,\;\forall d,\\
		&\mathbf{q}(d)(\mathbf{r}_n(d)+B)=0,\;\forall d.
	\end{align}
\end{subequations}

To obtain the optimal solution that satisfies these KKT conditions, we can first solve (\ref{r_n}) without constraints, i.e., $
		\textstyle\mathbf{r}_{n,\text{tmp}}^{k+1}=\frac{\max\{\|\hat{\mathbf{r}}_n^{k}\|_F-\tau^k\mu_x,0\}}{\|\hat{\mathbf{r}}_n^{k}\|_F}\hat{\mathbf{r}}_n^{k}$.
It is clear that if $\mathbf{r}_{n,\text{tmp}}^{k+1}$ satisfies the conditions in (\ref{r_n}), the problem is solved. 
However, if  $\mathbf{r}_{n,\text{tmp}}^{k+1}$ does not satisfy these conditions, we must consider different cases. 
We first define two index sets $\mathcal{S}_p\triangleq \{d:\;\mathbf{r}_{n,\text{tmp}}^{k+1}(d)>B\}$ and $\mathcal{S}_q\triangleq \{d:\;\mathbf{r}_{n,\text{tmp}}^{k+1}(d)<-B\}$ and consider
\begin{itemize}
	\item[1)] If $d\in\mathcal{S}_p$, then we should set $\mathbf{p}(d)>0$ and $\mathbf{q}(d)=0$ to reduce its absolute value and ensure the corresponding value of optimal vector  $\mathbf{r}_n^{k+1}(d)=B$.
	\item[2)] If $d\in\mathcal{S}_q$, then we should set $\mathbf{p}(d)=0$ and  $\mathbf{q}(d)>0$ to reduce its absolute value and ensure the corresponding value of optimal vector  $\mathbf{r}_n^{k+1}(d)=-B$.
	\item[3)] If $d\notin\mathcal{S}_p\cup\mathcal{S}_q$, then we should set $\mathbf{p}(d)=0$ and $\mathbf{q}(d)=0$ because positive $\mathbf{p}(d)$ and $\mathbf{q}(d)$ in cases 1) and 2) result in a smaller proximal coefficient $\frac{\max\left\{\left\|\hat{\mathbf{r}}_n^{k}-\mathbf{p}+\mathbf{q}\right\|_F\!-\tau^k\mu_x,0\right\}}{\left\|\hat{\mathbf{r}}_n^{k}-\mathbf{p}+\mathbf{q}\right\|_F}$, causing $|\mathbf{r}_n^{k+1}(d)|<|\mathbf{r}_{n,\text{tmp}}^{k+1}(d)|$. Consequently, $\mathbf{r}_n^{k+1}(d)$ would still satisfy the conditions.
\end{itemize}

The next step is to find the value of $\mathbf{p}$ and $\mathbf{q}$. Let $\mathbf{m}=\hat{\mathbf{r}}_n^{k}-\mathbf{p}+\mathbf{q}$, and  $a=\frac{\max\left\{\left\|\mathbf{m}\right\|_F-\tau^k\mu_x,0\right\}}{\left\|\mathbf{m}\right\|_F}$. Then, there are two cases:
\begin{itemize}
	\item[i)] If $\left\|\mathbf{m}\right\|_F>\tau^k\mu_x$, then $a=\frac{\left\|\mathbf{m}\right\|_F-\tau^k\mu_x}{\left\|\mathbf{m}\right\|_F}\in\!\left(0,1\right]$ and we have $\mathbf{r}_n^{k+1}=a\,\mathbf{m}$, i.e., $\mathbf{m}(d)=\hat{\mathbf{r}}_n^{k}(d)\,\mathbb{I}\{d\notin\mathcal{S}_p\cup\mathcal{S}_q\}+\frac{B}{a}\,\mathbb{I}\{d\in\mathcal{S}_p\}-\frac{B}{a}\,\mathbb{I}\{d\in\mathcal{S}_q\}$.
	Accordingly, $\left\|\mathbf{m}\right\|_F$ can be rewritten as $\textstyle\|\mathbf{m}\|_F=\textstyle\sqrt{\textstyle\frac{|\mathcal{S}_p\cup\mathcal{S}_q| B^2}{a^2}+\sum_{d\notin\mathcal{S}_p\cup\mathcal{S}_q}\hat{\mathbf{r}}_n^{k}(d)^2}$.
This can be substituted into $a=\frac{\left\|\mathbf{m}\right\|_F-\tau^k\mu_x}{\left\|\mathbf{m}\right\|_F}$ to obtain a quartic equation with respect to $a$.
Among the four solutions, the desired one lies within the range $(0,1]$.

	\item[ii)] If the aforementioned quartic equation has no solution in the range $(0,1]$, then we can only consider $\left\|\mathbf{m}\right\|_F\le \tau^k\mu_x$, i.e., $a=0$ and $\mathbf{r}_n^{k+1}=\mathbf{0}$.
	
\end{itemize}
After a maximum of $K$ iterations of the box-constrained FBS algorithm, we obtain $\mathbf{S}^{K+1}=[(\mathbf{H}^{K+1})^H, \mathbf{X}_{\text{D}}^{K+1}]^H\in\mathbb{C}^{(MP+R_D)\times N}$ and the active set of UEs is identified by comparing the UEs' channel energy to a threshold; expressly, we set $\hat{\xi}_n = 1$ if $\|\hat{\mathbf{h}}_n^{K+1}\|_F^2\ge T_{th}$ and $\hat{\xi}_n = 0$ otherwise.

\section{Simulation Results}
\subsection{Simulation Setup}
We consider a cell-free communication system containing $N=400$ uniformly distributed UEs  at a height of 1.65\,m, and $P$ uniformly distributed APs 
(we vary $P$ from $20$ to $100$) each with $M=4$ antennas and at a height 15\,m in an area of 500\,m $\times$ 500\,m.
We assume that the UE activity~$\{\xi_n\}_{\forall n}$ follows an i.i.d. Bernoulli distribution with $\mathbb{P}\{\xi_n=1\}=0.2,\;\forall n$. 
Each active UE transmits $R_p = 50$ pilots generated from a complex equiangular tight frame \cite{Tropp2005designing} with $\|\mathbf{x}_{\text{P},n}\|_F^2=R_P$, and $R_D = 200$ data signals with unit amplitude, i.e., $B=\sqrt{0.5}$, over the channel with a bandwidth of 20\,MHz and a carrier frequency of 1.9\,GHz. 
The transmit power of the UEs is 0.1\,W. We consider power control with a maximum dynamic power range of 12\,dB between the weakest and strongest UE.
Furthermore, shadow fading variance, noise figure, and noise temperature are 8\,dB, 9\,dB, and 290\,K, respectively.
For the FBS algorithms, we set the maximum number of iterations to $K=200$ and a stopping tolerance of $10^{-3}$.
In our simulations, we perform~$5 \cdot 10^3$ Monte--Carlo trials.
\subsection{Performance Metrics and Baseline Algorithms}
We consider the following performance metrics: user misdetection rate (UMR), channel estimation normalized mean square error (NMSE), average symbol error rate (ASER), and cumulative symbol error rate (CSER), which are defined as
\begin{subequations}
	\begin{align}
		&\textit{UMR}=\textstyle\frac{1}{N}{\sum_{n=1}^N|\xi_n-\hat{\xi}_n|},\\
		&\textit{NMSE}=\textstyle {\|\mathbf{H}-{\mathbf{H}}^{K+1}\|_F^2}/{\|\mathbf{H}\|_F^2},\\
		&\textit{ASER}=\textstyle\frac{1}{R_D N_a}{\sum_{n,r}\xi_n\mathbb{I}\{\mathbf{X}_\text{D}(n,r)\ne \hat{\mathbf{X}}_{\text{D}}^{K+1}(n,r)\}},\\
		&\textit{CSER}(x) =\textstyle \sum_{n_a = 1}^x \textit{ASER}(n_a)\mathbb{P}(N_a=n_a).
	\end{align}
\end{subequations}
Here, $\textit{ASER}(n_a)$ is the ASER for a specific number of active UEs $N_a=n_a$. CSER measures the impact of the number of active UEs on the symbol error rate.

To confirm the effectiveness of our algorithm, we compare it to different baselines, including ``\textit{Joint AUD-CE via \cite{goldstein2014field}, then DD},'' ``\textit{Joint AUD-CE via \cite{chen2018sparse}, then DD},'' and ``\textit{Joint AUD-CE-DD via \cite{song2022joint}}.'' Given the estimated channel matrix and active UEs via \cite{goldstein2014field} and \cite{chen2018sparse}, data detection is implemented by first performing zero-forcing equalization followed by mapping the result to the nearest QPSK symbol. To improve the convergence of JACD, we take the result of the baseline ``\textit{Joint AUD-CE via~\cite{goldstein2014field}, then DD}'' as the starting point for ``\textit{Joint AUD-CE-DD via \cite{song2022joint}}'' and our proposed FBS algorithm.

\begin{figure*}
	\centering
	\subfigure[Active user detection performance]{\label{AUD}\includegraphics[width=0.325\linewidth]{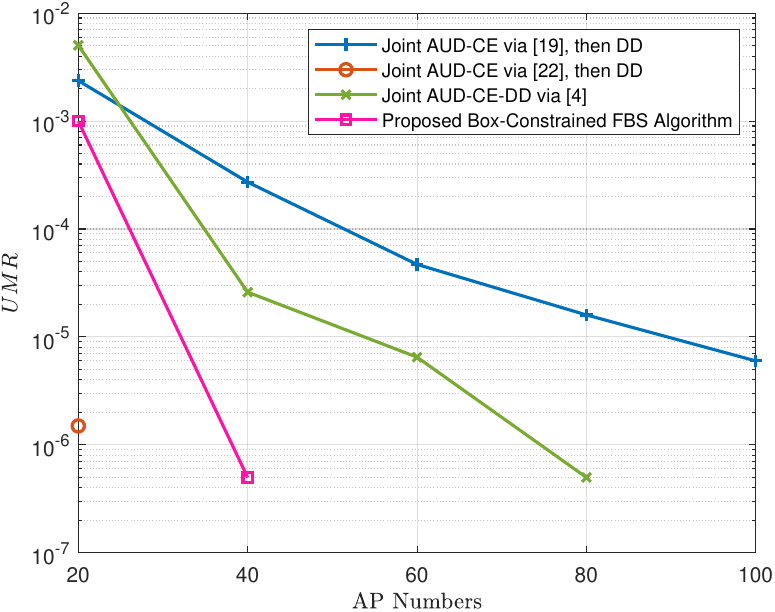}}
	\subfigure[Channel estimation performance]{\label{CE}\includegraphics[width=0.33\linewidth]{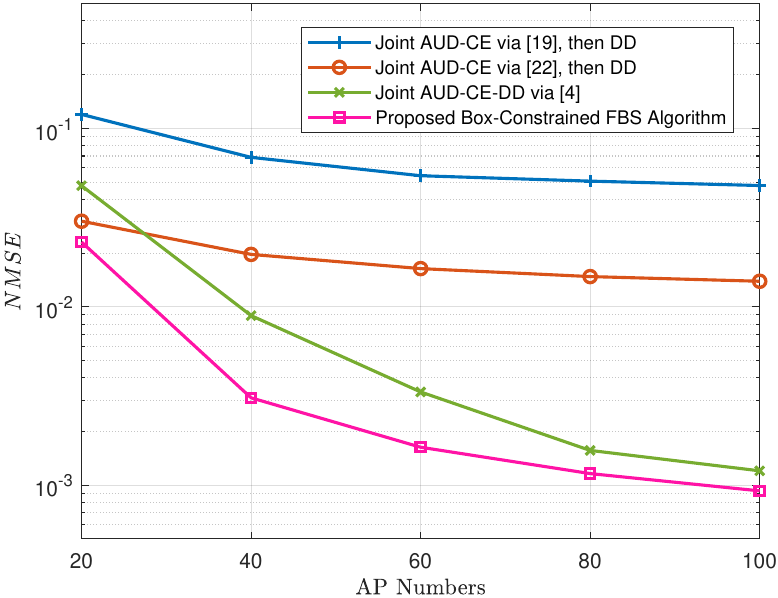}}
	\subfigure[Data detection performance]{\label{DD}\includegraphics[width=0.33\linewidth]{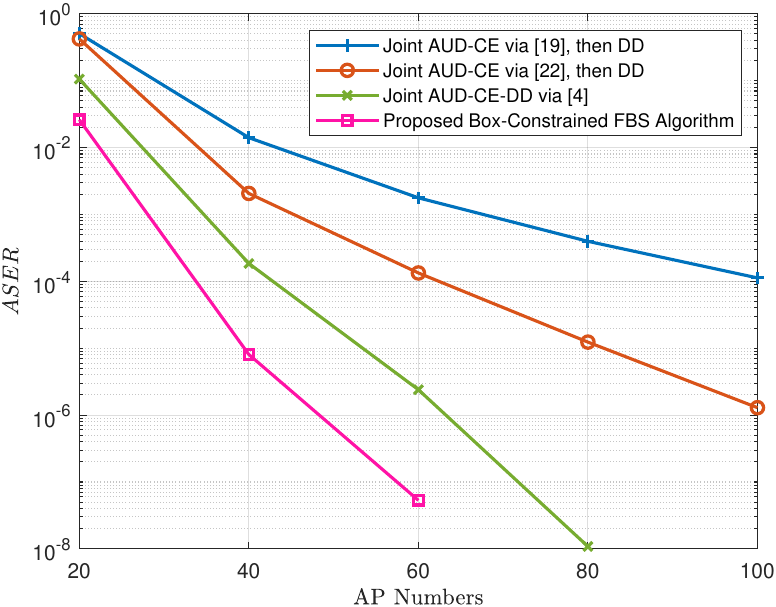}}
	\caption{Active user detection, channel estimation, and data detection performance comparison versus number of APs.}
	\label{main_fig}
	\vspace{-0.12cm}
\end{figure*}

\subsection{Simulation Results}
In Fig. \ref{main_fig}, we compare the different methods in terms of AUD, CE, and DD as the number of APs varies.
Our proposed algorithm outperforms all considered baseline methods in terms of AUD, CE, and DD in most scenarios.
Additionally, as the number of APs increases, the performance of all algorithms in AUD, CE, and DD improves and eventually stabilizes.
We also observe that our FBS algorithm can accurately detect the set of active UEs and their data when the number of APs is no less than $60$.

When comparing the ``Joint AUD-CE, then DD'' scheme, we note that the ``Joint AUD-CE-DD'' schemes generally outperform them in terms of DD performance.
This can be attributed to the received data signal containing implicit information about UEs' channels and activity, which helps to improve DD performance.
Moreover, compared with the ``\textit{Joint AUD-CE-DD via \cite{song2022joint}}'' baseline, we find that exploiting sparsity in the data matrix can improve JACD performance by providing additional UE sparse activity information to the proposed FBS algorithm.

\begin{figure}
	\centering
	\includegraphics[width=0.345\textwidth]{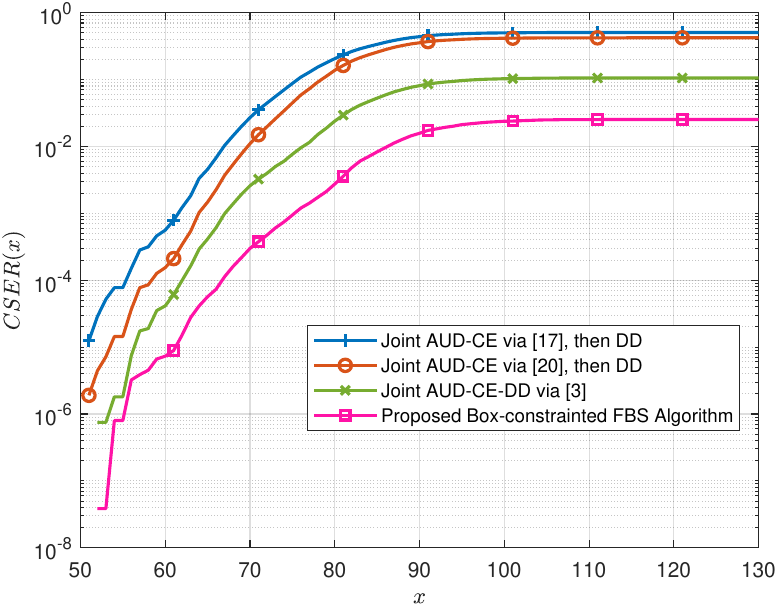}
	\vspace{-0.15cm}
	\caption{Cumulative symbol error rate comparison under $P=20$.}
	\label{Na}
\end{figure}

To illustrate the probability distribution of data detection errors for different numbers of active UEs $N_a$, we plot the CSER of different methods for $P=20$ in Fig.~\ref{Na}.
Our proposed algorithm exhibits the best CSER performance regardless of the value of $x$, followed by the ``\textit{Joint AUD-CE-DD via \cite{song2022joint}}'' baseline, which implies that our method achieves the best DD performance regardless of the number of active UEs.

\section{Conclusions}
We have proposed a novel framework for joint active user detection, channel estimation, and data detection for massive grant-free transmission in cell-free wireless communication systems. 
The proposed box-constrained forward-backward splitting algorithm outperforms existing methods in terms of active user detection, channel estimation, and data detection in most scenarios by exploiting the sparsity of both the cell-free channels and users' intermittent activity.

\balance

\bibliographystyle{IEEEtran}
\bibliography{./bib/Refabrv,./bib/IEEEBib1}

\begin{thebibliography}{10}
\providecommand{\url}[1]{#1}
\csname url@samestyle\endcsname
\providecommand{\newblock}{\relax}
\providecommand{\bibinfo}[2]{#2}
\providecommand{\BIBentrySTDinterwordspacing}{\spaceskip=0pt\relax}
\providecommand{\BIBentryALTinterwordstretchfactor}{4}
\providecommand{\BIBentryALTinterwordspacing}{\spaceskip=\fontdimen2\font plus
\BIBentryALTinterwordstretchfactor\fontdimen3\font minus
  \fontdimen4\font\relax}
\providecommand{\BIBforeignlanguage}[2]{{%
\expandafter\ifx\csname l@#1\endcsname\relax
\typeout{** WARNING: IEEEtran.bst: No hyphenation pattern has been}%
\typeout{** loaded for the language `#1'. Using the pattern for}%
\typeout{** the default language instead.}%
\else
\language=\csname l@#1\endcsname
\fi
#2}}
\providecommand{\BIBdecl}{\relax}
\BIBdecl

\bibitem{Sun2022massive}
G.~Sun, Y.~Li, X.~Yi, W.~Wang, X.~Gao, L.~Wang, F.~Wei, and Y.~Chen, ``Massive
  grant-free {OFDMA} with timing and frequency offsets,'' \emph{{IEEE} Trans.
  Wireless Commun.}, vol.~21, no.~5, pp. 3365--3380, May 2022.

\bibitem{Mishra2022rate}
A.~Mishra, Y.~Mao, L.~Sanguinetti, and B.~Clerckx, ``Rate-splitting assisted
  massive machine-type communications in cell-free massive {MIMO},''
  \emph{{IEEE} Commun. Lett.}, vol.~26, no.~6, pp. 1358--1362, Jun. 2022.

\bibitem{xu2023toward}
W.~Xu, Y.~Huang, W.~Wang, F.~Zhu, and X.~Ji, ``Toward ubiquitous and
  intelligent {6G} networks: from architecture to technology,'' \emph{Sci.
  China Inf. Sci.}, vol.~66, no.~3, p. 130300, Feb. 2023.

\bibitem{song2022joint}
H.~Song, T.~Goldstein, X.~You, C.~Zhang, O.~Tirkkonen, and C.~Studer, ``Joint
  channel estimation and data detection in cell-free massive {MU-MIMO}
  systems,'' \emph{{IEEE} Trans. Wireless Commun.}, vol.~21, no.~6, pp.
  4068--4084, Jun. 2022.

\bibitem{Ke2021massive}
M.~Ke, Z.~Gao, Y.~Wu, X.~Gao, and K.-K. Wong, ``Massive access in cell-free
  massive {MIMO}-based {Internet of Things}: Cloud computing and edge computing
  paradigms,'' \emph{{IEEE} J. Sel. Areas Commun.}, vol.~39, no.~3, pp.
  756--772, Mar. 2021.

\bibitem{xu2023edge}
W.~Xu, Z.~Yang, D.~W.~K. Ng, M.~Levorato, Y.~C. Eldar, and M.~Debbah, ``Edge
  learning for {B5G} networks with distributed signal processing: Semantic
  communication, edge computing, and wireless sensing,'' \emph{{IEEE} J. Sel.
  Topics Signal Process.}, vol.~17, no.~1, pp. 9--39, Jan. 2023.

\bibitem{Ganesan2021algorithm}
U.~K. Ganesan, E.~Björnson, and E.~G. Larsson, ``Clustering-based activity
  detection algorithms for grant-free random access in cell-free massive
  {MIMO},'' \emph{{IEEE} Trans. Wireless Commun.}, vol.~69, no.~11, pp.
  7520--7530, Nov. 2021.

\bibitem{Jiang2023EM}
S.~Jiang, J.~Dang, Z.~Zhang, L.~Wu, B.~Zhu, and L.~Wang, ``{EM-AMP}-based joint
  active user detection and channel estimation in cell-free system,''
  \emph{{IEEE} Syst. J.}, pp. 1--12, Early Access, 2023.

\bibitem{guo2022joint}
M.~Guo and M.~C. Gursoy, ``Joint activity detection and channel estimation in
  cell-free massive {MIMO} networks with massive connectivity,'' \emph{{IEEE}
  Trans. Commun.}, vol.~70, no.~1, pp. 317--331, Jan. 2022.

\bibitem{wang2022two}
X.~Wang, A.~Ashikhmin, Z.~Dong, and C.~Zhai, ``Two-stage channel estimation
  approach for cell-free {IoT} with massive random access,'' \emph{{IEEE} J.
  Sel. Areas Commun.}, vol.~40, no.~5, pp. 1428--1440, May 2022.

\bibitem{Iimori2021grant}
H.~Iimori, T.~Takahashi, K.~Ishibashi, G.~T.~F. de~Abreu, and W.~Yu,
  ``Grant-free access via bilinear inference for cell-free {MIMO} with
  low-coherence pilots,'' \emph{{IEEE} Trans. Wireless Commun.}, vol.~20,
  no.~11, pp. 7694--7710, Nov. 2021.

\bibitem{zou2020alow}
Q.~Zou, H.~Zhang, D.~Cai, and H.~Yang, ``A low-complexity joint user activity,
  channel and data estimation for grant-free massive {MIMO} systems,''
  \emph{{IEEE} Signal Process. Lett.}, vol.~27, pp. 1290--1294, Jul. 2020.

\bibitem{Di2022joint}
R.~B. Di~Renna and R.~C. de~Lamare, ``Joint channel estimation, activity
  detection and data decoding based on dynamic message-scheduling strategies
  for {mMTC},'' \emph{{IEEE} Trans. Commun.}, vol.~70, no.~4, pp. 2464--2479,
  Apr. 2022.

\bibitem{zhang2023joint}
S.~Zhang, Y.~Cui, and W.~Chen, ``Joint device activity detection, channel
  estimation and signal detection for massive grant-free access via {BiGAMP},''
  \emph{{IEEE} Trans. Signal Process.}, vol.~71, pp. 1200--1215, Apr. 2023.

\bibitem{jiang2020joint}
S.~Jiang, X.~Yuan, X.~Wang, C.~Xu, and W.~Yu, ``Joint user identification,
  channel estimation, and signal detection for grant-free {NOMA},''
  \emph{{IEEE} Trans. Wireless Commun.}, vol.~19, no.~10, pp. 6960--6976, Oct.
  2020.

\bibitem{sun2021OFDMA}
G.~Sun, Y.~Li, X.~Yi, W.~Wang, X.~Gao, and L.~Wang, ``{OFDMA} based massive
  grant-free transmission in the presence of timing offset,'' in \emph{Proc.
  13th Int. Conf. Wireless Commun. Signal Process. (WCSP)}, Changsha, China,
  Oct. 2021, pp. 1--6.

\bibitem{ngo2017cell}
H.~Q. Ngo, A.~Ashikhmin, H.~Yang, E.~G. Larsson, and T.~L. Marzetta,
  ``Cell-free massive {MIMO} versus small cells,'' \emph{{IEEE} Trans. Wireless
  Commun.}, vol.~16, no.~3, pp. 1834--1850, Mar. 2017.

\bibitem{Oscar2017bit}
O.~Castañeda, S.~Jacobsson, G.~Durisi, M.~Coldrey, T.~Goldstein, and
  C.~Studer, ``1-bit massive {MU-MIMO} precoding in {VLSI},'' \emph{IEEE J.
  Emerg. Sel. Topics Circuits Syst.}, vol.~7, no.~4, pp. 508--522, Dec. 2017.

\bibitem{goldstein2014field}
T.~Goldstein, C.~Studer, and R.~Baraniuk, ``A field guide to forward-backward
  splitting with a {FASTA} implementation,'' \emph{arXiv preprint
  arXiv:1411.3406}, Nov. 2014.

\bibitem{beck2009fast}
A.~Beck and M.~Teboulle, ``A fast iterative shrinkage-thresholding algorithm
  for linear inverse problems,'' \emph{SIAM J. Imag. Sci.}, vol.~2, no.~1, pp.
  183--202, 2009.

\bibitem{Tropp2005designing}
J.~Tropp, I.~Dhillon, R.~Heath, and T.~Strohmer, ``Designing structured tight
  frames via an alternating projection method,'' \emph{{IEEE} Trans. Inf.
  Theory}, vol.~51, no.~1, pp. 188--209, Jan. 2005.

\bibitem{chen2018sparse}
Z.~Chen, F.~Sohrabi, and W.~Yu, ``Sparse activity detection for massive
  connectivity,'' \emph{{IEEE} Trans. Signal Process.}, vol.~66, no.~7, pp.
  1890--1904, Apr. 2018.

\end{thebibliography}

\end{document}